\documentclass{llncs}

\usepackage[utf8]{inputenc}
\usepackage{fontenc}
\usepackage{url}
\usepackage{textcomp}
\usepackage{listings}
\usepackage{graphicx}

\newenvironment{squote}{\begin{quote}\footnotesize}{\normalsize\end{quote}}

\begin{document}

\title{Towards a Study of Meta-Predicate Semantics}
\titlerunning{Towards a Study of Meta-Predicate Semantics}

\author{Paulo Moura}
\authorrunning{Paulo Moura}

\institute{
	Dep. of Computer Science, University of Beira Interior, Portugal\\
	Center for Research in Advanced Computing Systems, INESC--Porto, Portugal\\
	\email{pmoura@di.ubi.pt}
}

\maketitle

\begin{abstract}
We describe and compare design choices for meta-predicate semantics, as found in representative Prolog module systems and in Logtalk. We look at the consequences of these design choices from a pragmatic perspective, discussing  explicit qualification semantics, computational reflection support, expressiveness of meta-predicate declarations, safety of meta-predicate definitions, portability of meta-predicate definitions, and meta-predicate performance. Our aim is to provide useful insight for debating meta-predicate semantics and portability issues based on actual implementations and common usage patterns.\\


\noindent
\textbf{Keywords:} meta-predicate semantics, modules, objects.
\end{abstract}

\section{Introduction}
\label{intro}

Prolog and Logtalk \cite{pmoura03,lgtuserman2390} meta-predicates are predicates with one or more arguments that are either goals or closures\footnote{In Prolog and Logtalk, a closure is defined as a callable term used to construct a goal by appending one or more additional arguments.} used for constructing goals, which are called in the body of a predicate clause. Common examples are all-solutions meta-predicates such as \lstinline{setof/3} and list mapping predicates. Prolog implementations may also classify predicates as meta-predicates whenever the predicate arguments need to be module-aware. Examples include the built-in database predicates such as \lstinline{assertz/1} and \lstinline{retract/1}.

Meta-predicates provide a mechanism for reusing programming patterns. By encapsulating meta-predicate definitions in modules and objects, exported and public meta-predicates allow client modules and client objects to reuse these patterns, customized by calls to local predicates.

In order to compare meta-predicate implementations, a number of design choices can be considered. These include explicit qualification semantics, computational reflection support, expressiveness of meta-predicate declarations, safety of meta-predicate definitions, portability of meta-predicate definitions, and meta-predicate performance.

When discussing meta-predicate semantics, it is useful to define the contexts where a meta-predicate is defined, called, and executed. The following definitions are taken from \cite{pmoura09b} and will be used in this paper:

\begin{description}
	\item[Definition context] This is the object or module containing the meta-predicate definition.

	\item[Calling context] This is the object or module from which a meta-predicate is called. This can be the object or module where the meta-predicate is defined in the case of a local call or another object or module assuming that the meta-predicate is within scope.

	\item[Execution context] This includes both the calling context and the definition context. It is comprised by all the information necessary for the language runtime to correctly execute a meta-predicate call.
\end{description}

\noindent
In this paper, we make use of an additional definition:

\begin{description}
	\item[Lookup context] This is the object or module where we start looking for the meta-predicate definition. The definition can always be reexported from another module or inherited from another object.
\end{description}


This paper is organized as follows. Section 2 describes meta-predicate directives. Section 3 discusses the consequences of using explicit qualified meta-predicate calls and the transparency of control constructs when using explicit qualification. Section 4 describes the support for computational reflection for meta-predicates on Logtalk and Prolog module systems. Section 5 briefly presents a set of compilation safety rules for meta-predicate definitions. Section 6 discusses the portability of meta-predicate directives and meta-predicate definitions. Section 7 presents some remarks on meta-predicate performance. Section 8 sumarizes our conclusions.

\section{Meta-Predicate Directives}
\label{directive}

Meta-predicate directives are required for proper compilation of meta-predicates in both Logtalk and Prolog module systems. The design choices behind the current variations of meta-predicates directives translate to different trade-offs between simplicity and expressiveness.

\subsection{The ISO Prolog Standard \lstinline{metapredicate/1} Directive}

The ISO Prolog standard for Modules \cite{isoiec00} specifies a \lstinline{metapredicate/1} directive that allows us to describe which meta-predicate arguments are normal arguments and which are meta-arguments using a predicate template. In this template, the atom \lstinline{*} represents a normal argument while the atom \lstinline{:} represents a meta-argument. We are not aware of any Prolog module system implementing this directive. The standard does allow for alternative meta-predicate directives, providing solely as an example a \lstinline{meta/1} directive that takes a predicate indicator as argument. This alternative directive seems familiar to the \lstinline{tool/2} and \lstinline{module_transparent/1} directives discussed below. However, from the point-of-view of standardization and code portability, allowing for alternative directives is more harmful than helpful.

\subsection{The Prolog \lstinline{meta_predicate/1} Directive}

The ISO Prolog specification of a meta-predicate directive suffers from major shortcomings \cite{richard}. First, it is not possible to distinguish between goals and closures. Second, it is not possible to represent the instantiation mode of the normal arguments (the instantiation mode of meta-arguments is implicit).

The de facto standard solution for specifying closures is to use a non-negative integer representing the required number of additional arguments \cite{quintusman35}. By interpreting a goal as a closure requiring zero additional arguments, we can reserve the atom \lstinline{:} to represent arguments that need to be module (or object) aware without necessarily referring to a predicate. This convention is found in recent SICStus Prolog \cite{sicstususerman41} and SWI-Prolog versions \cite{swiman5910} and is being adopted by other Prolog compilers. In Prolog module systems where module expansion only needs to distinguish between normal arguments and meta-arguments, using an integer for representing closures can still be useful for cross-reference tools.

For representing the instantiation mode of normal arguments, the atoms \lstinline{+}, \lstinline{?}, \lstinline{@}, and \lstinline{-} are commonly used, as specified in the ISO Prolog standard \cite{isoiec95}.

Despite the level of detail in the description of meta-predicate arguments, there is, however, a known representation shortcoming. Some predicates accept a list of options where one or more options are module-aware. For example, the third argument of the predicate \lstinline{thread_create/3} \cite{mtISOdraft} is a list of options that can include an \lstinline{at_exit/1} option. This option specifies a goal to be executed when a thread terminates. In this case, the argument is not a meta-argument but may \textsl{contain} a sub-term that will be used as a meta-argument. Although we could devise (a most likely cumbersome) syntax for these cases, the elegant solution for this representation problem is provided by the \lstinline{tool/2} and \lstinline{module_transparent/1} directives discussed below.

\subsection{The Logtalk \lstinline{meta_predicate/1} Directive}

Logtalk uses a \lstinline{meta_predicate/1} directive with the atom \lstinline{:} replaced by \lstinline{::} for consistency with the message sending operator and allowing \lstinline{::} to be synonym to the integer zero to indicate goal arguments. As in the Prolog directive described above, closures are represented by a non-negative integer. Logtalk uses this information to verify meta-predicate definitions, as discussed in \cite{pmoura09b}. Logtalk supports a \lstinline{mode/2} predicate directive for specifying the instantiation mode and the type of predicate arguments (plus the predicate determinism). Therefore, the atom \lstinline{*} is used to indicate normal arguments in \lstinline{meta_predicate/1} directives.

\subsection{The \lstinline{tool/2} and \lstinline{module_transparent/1} Directives}

An alternative, used in ECLiPSe \cite{eclipseman03} and in earlier SWI-Prolog versions \cite{Wielemaker:03b} is to simply declare meta-predicates as \textsl{module transparent}, forgoing the specification of which arguments are normal arguments and which arguments are meta-arguments. For this purpose, ECLiPSe provides a \lstinline{tool/2} directive and SWI-Prolog provides a (now deprecated) \lstinline{module_transparent/1} directive. These directives take predicate indicators as arguments and thus support a simpler and user-friendlier solution when compared with the \lstinline{meta_predicate/1} directive. However, we have show in \cite{pmoura09b} that distinguishing between goals and closures and specifying the exact number of closure additional arguments is necessary to avoid misuse of meta-predicate definitions.

\section{Explicit Qualification Semantics}
\label{explicit}

The semantics of explicit qualification is the single most significant design decision on meta-predicate semantics. This section compares two different semantics, found on actual implementations, for the explicit qualification of meta-predicate and control constructs.

\subsection{Explicit Qualification of Meta-Predicate Calls}

Given an explicit qualified meta-predicate call, we have two sensible choices for the corresponding semantics:

\begin{enumerate}
	\item The explicit qualification sets only the initial lookup context for the meta-predicate definition. Therefore, all meta-arguments that are not explicitly-qualified are called in the meta-predicate calling context.\\
	\item The explicit qualification sets both the initial lookup context for the meta-predicate definition and the meta-predicate calling context. Therefore, all meta-arguments that are not explicitly-qualified are called in the meta-predicate lookup context (usually the same as the meta-predicate definition context).
\end{enumerate}

These two choices for explicit qualification semantics are also described in the ISO Prolog standard for modules. This standard specifies a read-only flag, \lstinline{colon_sets_calling_context}, which would allow a programmer to query the semantics of a particular module implementation.

Logtalk and the ECLiPSe module system implement the first choice. Prolog module systems derived from the Quintus Prolog module system \cite{quintusman35}, including those found on Ciao Prolog, SICStus Prolog, SWI-Prolog, and YAP implement the second choice (the native XSB module system is atom-based, not predicate-based; we will not discuss it here).



In order to illustrate the differences between the two choices above, consider the following example, running on Prolog module systems implementing the second choice. First, we define a meta-predicate library:

\begin{squote}
\begin{verbatim}
:- module(library, [my_call/1]).

:- meta_predicate(my_call(0)).
my_call(Goal) :-
    write('Calling: '), writeq(Goal), nl, call(Goal).

me(library).
\end{verbatim}
\end{squote}

\noindent
Second, we define a simple client module:

\begin{squote}
\begin{verbatim}
:- module(client, [test/1]).

:- use_module(library, [my_call/1]).

test(Me) :-
    my_call(me(Me)).

me(client).
\end{verbatim}
\end{squote}

\noindent
To test our code, we use the following query:

\begin{squote}
\begin{verbatim}
?- client:test(Me).

Calling: client:me(_)
Me = client
yes
\end{verbatim}
\end{squote}

\noindent
This query provides the expected result. But consider the following seemingly innocuous changes to the client module:

\begin{squote}
\begin{verbatim}
    :- module(client, [test/1]).

    test(Me) :-
        library:my_call(me(Me)).

    me(client).
\end{verbatim}
\end{squote}

\noindent
In this second version, we use explicit qualification in order to call the \lstinline{my_goal/1} meta-predicate. Repeating our test query gives:

\begin{squote}
\begin{verbatim}
?- client:test(Me).

Calling: library:me(_)
Me = library
yes
\end{verbatim}
\end{squote}

\noindent
In order for a programmer to understand this result, we needs to be aware that the \lstinline{:/2} operator both calls a predicate in another module and changes the calling context of the predicate to that module. The first use is expected. The second use is not obvious, is counterintuitive (due to different semantics between implicitly-qualified and explicitly-qualified calls to the same predicate), and often not properly documented. In the common case where we are reusing a library meta-predicate, the user (rightfully) expects that a local predicate will be called when using a meta-argument with the same functor and arity instead of a predicate with the same name in the meta-predicate definition context. Indeed, is unlikely that a predicate with the same name of the local predicate even exist in the library module. We can, however, conclude that the meta-predicate definition is still working as expected as the calling context is set to the library module. If we still want the \lstinline{me/1} predicate to be called in the context of the client module instead, we need to explicitly qualify the meta-argument by writing:

\begin{squote}
\begin{verbatim}
test(Me) :-
    library:my_call(client:me(Me)).
\end{verbatim}
\end{squote}

This is an awkward solution but it works as expected in the rare cases that require explicit qualification. It should be noted that the idea of the \lstinline{meta_predicate/1} directive is to avoid the need for explicit qualifications in the first place. But that requires using \lstinline{use_module/1-2} directives for importing the meta-predicates and implicit qualification when calling them. This explicit qualification of meta-arguments is not necessary in Logtalk and in the ECLiPSe module system, where explicit qualification of a meta-predicate call sets where to start looking for the meta-predicate definition, not where to look for the meta-arguments definition.

The semantics of the \lstinline{:/2} operator in Prolog module systems derived from the Quintus Prolog module system is probably rooted in optimization goals. When a directive \lstinline{use_module/1} is used, most (if not all) Prolog compilers require the definition of the imported module to be available (thus resolving the call at compilation time). However, that does not seem to be required when compiling an explicitly qualified module call. For example, using recent versions of SICStus Prolog, SWI-Prolog, and YAP, the following code compiles without errors or warnings (despite the fact that the module \lstinline{fictitious} does not exist):

\begin{squote}
\begin{verbatim}
:- module(client, [test/1]).

test(X) :-
    fictitious:predicate(X).
\end{verbatim}
\end{squote}

Thus, in this case the \lstinline{fictitious:predicate/1} call is resolved at runtime. In our example above with the explicit call to the \lstinline{my_call/1} meta-predicate, the implementation of the \lstinline{:/2} operator propagates at runtime the module prefix to the meta-arguments that are not explicitly qualified. This runtime propagation translates to a performance penalty. Therefore, and not surprisingly, the use of explicit qualification is discouraged by the Prolog implementers. In fact, until recently, most Prolog implementations provided poor performance for \lstinline{:/2} calls even when the necessary module information was available at compile time.


\subsection{Transparency of Control Constructs}

One of the design choices regarding meta-predicate semantics is the transparency of control constructs to explicit qualification. The relevance of this topic is that most control constructs can also be regarded as meta-predicates. In fact, there is a lack of agreement on the Prolog community on which language elements are control constructs and which language elements are predicates. For the purposes of our discussion, we use the classification found on the ISO Prolog standard, which specifies the following control constructs: \lstinline{call/1}, \textsl{conjunction}, \textsl{disjunction}, \textsl{if-then}, \textsl{if-then-else}, and \lstinline{catch/3}. The standard also specifies \lstinline{true/0}, \lstinline{fail/0}, \lstinline{!/0}, and \lstinline{throw/1} as control constructs but none of these can be interpreted as a meta-predicate.

When a control construct is transparent to explicit qualification, the qualification propagates to all the control constructs arguments that are not explicitly qualified. For example, the following equivalences hold:\\

\begin{tabular}{lcl}
\lstinline!M:(A, B)! & $\Leftrightarrow$ & \lstinline!(M:A, M:B)! \\
\lstinline!M:(A; B)! & $\Leftrightarrow$ & \lstinline!(M:A; M:B)! \\
\lstinline!M:(A -> B; C)! & $\Leftrightarrow$ & \lstinline!(M:A -> M:B; M:C)! \\
\end{tabular}

\vspace{1em}

\noindent
In Prolog module systems where the \lstinline{:/1} operator sets both the meta-predicate lookup context and the meta-arguments calling context, the above equivalences are consistent with the explicit qualification semantics of meta-predicates described in the previous section. For example:\\

\begin{tabular}{lcl}
\lstinline!M:findall(T, G, L)! & $\Leftrightarrow$ & \lstinline!findall(T, M:G, L)! \\
\lstinline!M:assertz(A)! & $\Leftrightarrow$ & \lstinline!assertz(M:A)! \\
\end{tabular}
\vspace{1em}

\noindent
This is also true for user-defined meta-predicates. For the example presented in the previous section, the following equivalence holds:\\

\begin{tabular}{lcl}
\lstinline!library:my_call(me(Me))! & $\Leftrightarrow$ & \lstinline!my_call(library:me(Me))! \\
\end{tabular}

\vspace{1em}

\noindent
We can conclude that the different semantics of implicitly and explicitly qualified meta-predicate calls, which is at odds with most user expectations, allows the semantics of explicitly qualified control constructs to be consistent with the semantics of explicitly qualified meta-predicate calls.

In systems such as ECLiPSe or Logtalk, where explicit qualification only sets the lookup context, the semantics of control constructs and meta-predicates are different. In Logtalk, the following equivalences for control constructs are handy, supported, and can be interpreted as a shorthand notation for sending a set of messages to the same object (the \lstinline{::/2} operator is used in Logtalk for message sending):\\

\begin{tabular}{lcl}
\lstinline!O::(A, B)! & $\Leftrightarrow$ & \lstinline!(O::A, O::B)! \\
\lstinline!O::(A; B)! & $\Leftrightarrow$ & \lstinline!(O::A; O::B)! \\
\lstinline!O::(A -> B; C)! & $\Leftrightarrow$ & \lstinline!(O::A -> O::B; O::C)! \\
\end{tabular}

\vspace{1em}

\noindent
ECLiPSe implements a simpler design choice, disallowing the above shorthands, and thus treating control constructs and meta-predicates uniformly. Both Logtalk and ECLiPSe provide, however, the same syntax for implicitly and explicitly qualified meta-predicate calls. Consider the following objects, corresponding to a a Logtalk version of the Prolog module example used in the previous section: 

\begin{squote}
\begin{verbatim}
:- object(library).

    :- public(my_call/1).
    :- meta_predicate(my_call(::)).
    my_call(Goal) :-
        write('Calling: '), writeq(Goal), nl,
        call(Goal),
        sender(Sender), write('Sender:  '), writeq(Sender).

    me(library).

:- end_object.


:- object(client).

    :- public(test/1).
    test(Me) :-
        library::my_call(me(Me)).

    me(client).

:- end_object.
\end{verbatim}
\end{squote}

\noindent
Our test query becomes:

\begin{squote}
\begin{verbatim}
?- client::test(Me).

Calling: me(_G216)
Sender:  client
Me = client.
yes
\end{verbatim}
\end{squote}

\noindent
That is, meta-arguments are always called in the context of the meta-predicate call. Logtalk also implements common built-in meta-predicates such as \lstinline{call/1-N}, \lstinline{\+/1}, \lstinline{findall/3}, and \lstinline{phrase/3} with the same semantics as user-defined meta-predicates. In order to avoid misinterpretations, these built-in meta-predicates are implemented as private predicates.\footnote{Logtalk supports \textsl{private}, \textsl{protected}, and \textsl{public} predicates. A predicates may also be \textsl{local} if no scope directive is present, making the predicate invisible to the built-in reflection predicates.} Thus, the following call is illegal and results in a permission error:

\begin{squote}
\begin{verbatim}
?- some_object::findall(T, g(T), L).

error(
  permission_error(access,private_predicate,findall(T,g(T),L)),
  some_object::findall(T,g(T),L),
  user)
\end{verbatim}
\end{squote}

\noindent
The correct call would be:

\begin{squote}
\begin{verbatim}
?- findall(T, some_object::g(T), L).
\end{verbatim}
\end{squote}

\noindent
We can conclude that ensuring the same semantics for implicitly and explicitly qualified meta-predicate calls requires different semantics for explicitly qualified control constructs, and thus a clear distinction between control constructs and predicates, in order to provide a reading for explicitly qualified control constructs that matches user expectations. This distinction can be rendered moot, however, if we simply disallow explicit qualification of control constructs, as exemplified by ECLiPSe.

\section{Computational Reflection Support}
\label{reflection}

Computational reflection allows us to perform computations about the \textsl{structure} and the \textsl{behavior} of an application. In the case of meta-predicates, structural reflection allows us to find where the meta-predicate is defined and about the meta-predicate template, while behavioral reflection allows us to access the meta-predicate execution context. As described in Section \ref{intro}, a meta-predicate execution context includes information about from where the meta-predicate is called. This is only meaningful, however, in the presence of a predicate encapsulation mechanism such as modules or objects. Access to the execution-context is usually not required for common user-level meta-predicate definitions but can be necessary when meta-predicates are used, for example, to extend Prolog or Logtalk meta-call features. In the case of Logtalk, full access to predicate execution context is provided by the \lstinline{sender/1}, \lstinline{self/1}, \lstinline{this/1}, and \lstinline{parameter/2} built-in predicates \cite{lgtuserman2390}. For Prolog compilers supporting modules, the following table provides an overview of the reflection built-in predicates that can be used to access a meta-predicate execution context:

	\begin{center}
		\begin{tabular}{ccc}
			\hline
			\textbf{Prolog compiler}    & \textbf{Built-in reflection predicates} \\
			\hline
			Ciao 1.10          & \lstinline!predicate_property/2! (in library \lstinline!prolog_sys!)\\
			ECLiPSe 6.0        & \lstinline!get_flag/3! \\
			SICStus Prolog 4.1 & \lstinline!predicate_property/2! \\
			SWI-Prolog 5.9.10  & \lstinline!context_module/1!, \lstinline!predicate_property/2!, \lstinline!strip_module/3! \\
			XSB 3.2            & \lstinline!predicate_property/2! \\
			YAP 6.0            & \lstinline!context_module/1!, \lstinline!predicate_property/2! \\
			\hline
		\end{tabular}
	\end{center}


From this table we conclude that the only built-in predicate common to these Prolog compilers is \lstinline{predicate_property/2}. Together with the ECLiPSe \lstinline{get_flag/3} and the SWI-Prolog and YAP \lstinline{context_module/1} predicates, these built-ins only provide \textsl{structural} reflection. Specifically, information about the meta-predicate template and the definition context of the meta-predicate. SWI-Prolog is the only compiler that provides \textsl{built-in} access to the meta-predicate calling context using the predicate \lstinline{strip_module/3}. As a simple example of using this predicate consider the following module:
\begin{squote}
\begin{verbatim}
:- module(m, [mp/2]).

:- meta_predicate(mp(0, -)).

mp(Goal, Caller) :-
    strip_module(Goal, Caller, _),
    call(Goal).
\end{verbatim}
\end{squote}

\noindent
After compiling and loading this module, the following queries illustrate both the functionality of the \lstinline{strip_module/3} predicate and the consequences of explicit qualification of the meta-predicate call:

\begin{squote}
\begin{verbatim}
?- mp(true, Caller).
Caller = user.

?- m:mp(true, Caller).
Caller = m.
\end{verbatim}
\end{squote}

\noindent
For similar Prolog compiler module systems, descending from the Quintus Prolog module system, it is possible to access the meta-predicate calling context by looking into the implicit qualification of a meta-argument: 
\begin{squote}
\begin{verbatim}
:- module(m, [mp/2]).

:- meta_predicate(mp(0, -)).

mp(Goal, Caller) :-
    Goal = Caller:_,
    call(Goal).
\end{verbatim}
\end{squote}

\noindent
After compiling and loading this module, we can reproduce the results illustrated by the queries above for the SWI-Prolog version of this module. The only possible caveat would be if the Prolog compiler fails to ensure that there is always a single qualifier for a goal. That is, that terms such as \lstinline{M1:(M2:(M3:G))} are never generated internally when propagating module qualifications.\\

In the case of ECLiPSe, a built-in predicate for accessing the meta-predicate calling context is not necessary as the \lstinline{tool/2} directive works by connecting a meta-predicate interface with its implementation:
\begin{squote}
\begin{verbatim}
:- module(m).

:- export(mp/2).
:- tool(mp/2, mp/3).

mp(Goal, Caller, Caller) :-
    call(Goal).
\end{verbatim}
\end{squote}

\noindent
After compiling and loading this module, repeating the above queries illustrate the difference in explicit qualification semantics between ECLiPSe and the other Prolog compilers:
\begin{squote}
\begin{verbatim}
[eclipse 16]: mp(true, Caller).

Caller = eclipse
Yes (0.00s cpu)
[eclipse 17]: m:mp(true, Caller).

Caller = eclipse
Yes (0.00s cpu)
\end{verbatim}
\end{squote}

\noindent
Note that the module \lstinline{eclipse} is the equivalent of the module \lstinline{user} in other compilers.

\section{Secure Meta-Predicate Definitions}
\label{secure}

Meta-predicate definitions should not provide a mechanism for calling client predicates other than the ones intended by the meta-predicate calls. This, however, is mostly meaningful for languages such as Logtalk and for Prolog module systems, such as ECLiPSe and Ciao \cite{ciaomodules00}, that aim to enforce object and module predicate scope rules. The following set of compilation rules, described and illustrated in detail in \cite{pmoura09b}, contribute to make meta-predicate definitions secure:

\begin{enumerate}
	\item The meta-arguments of a meta-predicate clause head must be variables.\\
	\item Meta-calls whose arguments are not variables appearing in meta-argument positions in the clause head must be compiled as calls to local predicates.\\
	\item Meta-predicate closures must be used within a \lstinline{call/2-N} built-in predicate call that complies with the corresponding meta-predicate directive.
\end{enumerate}

These rules are implemented in Logtalk. For Prolog module systems whose design allows any module predicate to be called using explicit module qualification, these rules may be regarded as best practice for writing meta-predicates and thus useful for checking meta-predicate definitions for possible errors (e.g. as part of lint checkers).

\section{Meta-predicate Definitions Portability}

The portability of meta-predicate definitions depends on two main factors: portability of the meta-predicate directives and portability of the meta-call primitives used when implementing the meta-predicates. Other factors that may impact portability are the preprocessing solutions for improving meta-predicate performance, described in Section \ref{performance}, and the mechanisms for computational reflection about meta-predicate definition and execution, discussed in Section \ref{reflection}.

\subsection{Specification of Closures and Instantiation Modes in Meta-Predicate Directives}

The main portability issue of meta-predicate directives is the use of non-negative integers to specify closures and the atoms used to specify the instantiation mode of normal arguments.

Although the use of non-negative integers comes from Quintus Prolog, it was mostly regarded as a way to provide information to cross-reference and documentation tools. Prolog compilers such as SICStus Prolog \cite{sicstususerman41} and YAP \cite{yapdocs} accept this notation but only for compatibility with existing code. Other Prolog compilers such as Ciao define alternative but incompatible syntaxes for specifying closures.

There is also some variation in the atoms used for representing the instantiation modes of normal arguments. Some Prolog compilers use an extended set of atoms for documenting argument instantiation modes compared to the basic set (\lstinline{+}, \lstinline{?}, \lstinline{@}, and \lstinline{-}) found, for example, in the ISO Prolog standard. It is therefore tempting to use these extended sets in meta-predicate directives, which will likely raise portability issues.

We hope that recent Prolog standardization initiatives, specially the development of portable libraries, will lead to establish a de facto standard meta-predicate directive derived from the extended directive described in Section \ref{directive}.

\subsection{The call/1-N Control Constructs}

The \lstinline{call/1} control construct is specified in the ISO Prolog standard \cite{isoiec95}. This control construct is implemented by virtually all Prolog compilers. The \lstinline{call/2-N} control constructs, whose use is strongly recommended whenever a meta-predicate works with closures (see Section \ref{secure}), is specified in the ISO Prolog standard Core Revision proposal \cite{crISOdraft}.
 A growing number of Prolog compilers implement these control constructs but with different maximum values for \lstinline{N}, which can raise some portability problems. Ideally, the implementation of the \lstinline{call/1-N} control constructs would support \lstinline{N} up to the maximum predicate arity. However, the feasibility of supporting a large value for \lstinline{N} depends on the design decisions of a Prolog compiler implementation. It should also be noted that, in some Prolog compilers such as recent SWI-Prolog and YAP versions, the maximum predicate arity of a Prolog compiler is unbounded. From a language design point-of-view, limiting the maximum value of \lstinline{N} to a value different from maximum term arity can be interpreted as a flaw. It certainly seems odd that a programmer can use the \lstinline{=../2} built-in predicate and the \lstinline{call/1} control construct to build and call a goal but that a similar solution cannot be used by the Prolog implementer when compiling \lstinline{call/2-N} calls. From a pragmatic point-of-view, it is unlikely that user written code (not necessarily user \textsl{generated} code) would require a large upper limit of \lstinline{N}. Despite the apparent lack of agreement, the more significant portability issues here are Prolog compilers only supporting \lstinline{call/1} or a arguably small value of \lstinline{N}. The following table summarizes the implementations of the \lstinline{call/2-N} control construct on selected Prolog compilers:

\vspace{-0.5em}

	\begin{center}
		\begin{tabular}{ccc}
			\hline
			\textbf{System}    & \textbf{N}   & \textbf{Notes} \\
			\hline
			B-Prolog 7.4       & 10/65535 & (interpreter/compiler i.e. maximum arity) \\
			Ciao 1.10          & 255      & (maximum arity using the \lstinline!hiord! library) \\
			CxProlog 0.94.0    & 9        & --- \\
			ECLiPSe 6.0        & 1        & --- \\
			GNU Prolog 1.3.1   & 11       & --- \\
			JIProlog 3.0.2     & 5        & --- \\
			K-Prolog 6.0.4     & 9        & --- \\
			Qu-Prolog 8.10     & 1        & (supports a \lstinline!call_predicate/1-5! built-in predicate) \\
			SICStus Prolog 4.1 & 255      & (maximum arity) \\
			SWI-Prolog 5.9.10  & 8        & (\lstinline!meta_predicate/1! directive limit) \\
			XSB 3.2            & 11       & --- \\
			YAP 6.0            & 12       & --- \\
			\hline
		\end{tabular}
	\end{center}

This table only lists \textsl{built-in} support for \lstinline{call/2-N} control construct. While this control construct can be defined by the programmer using the built-in predicate \lstinline{=../2} and an \lstinline{append/3} predicate, such definitions provide relative poor performance due to the construction and appending of temporary lists of arguments.

\section{Meta-Predicate Performance}
\label{performance}

Considering that meta-programming is often touted as a major feature of Prolog, the relative poor performance of meta-calls is embarrassing. For those cases where performance is an important factor, the usual solution is to interpret meta-predicate definitions as high-level macros and to preprocess meta-predicate calls in order to replace them with calls to auxiliary predicate definitions that do not contain meta-calls. This preprocessing is usually only performed on code marked as stable as the auxiliary predicates often complicate debugging. The preprocessing code is often implemented in optional libraries, which can be found on several Prolog compilers such as ECLiPSe, SWI-Prolog, and YAP. These libraries, however, require custom code for each meta-predicate. Therefore, user-defined meta-predicates will fail to match the performance of library-supported meta-predicates unless the user also writes its own custom preprocessing code. A more generic solution for preprocessing meta-predicate definitions is needed to make using these programming patterns more appealing for performance-critical applications.

\section{Conclusions}

We presented a comprehensive set of meta-predicate design decisions based on current practice in 
Logtalk and Prolog module systems. The most remarkable result is that none of the two commonly implemented semantics for explicitly qualified calls provides an ideal solution that both matches user expectations and allows the distinction between predicates and control constructs to be waived. By describing the consequences of these design decisions we provided useful insight to discuss meta-predicate semantics, often a difficult subject for inexperienced programmers. We hope that this paper contributes to a convergence of meta-predicate directive syntax, meta-predicate semantics, and meta-predicate related reflection built-in predicates among Prolog module systems.

\subsubsection*{Acknowledgements.} We are grateful to Ulrich Neumerkel, Jan Wielemaker, and Richard O'Keefe for interesting discussions about explicitly-qualified meta-predicate call semantics on the SWI-Prolog mailing list. We thank also the anonymous reviewers for their informative comments. This work is partially supported by FCT project MOGGY -- PTDC/EIA/70830/2006.

\bibliographystyle{splncs}
\bibliography{semantics}

\end{document}